\documentclass[manuscript,screen]{acmart}
\usepackage{longtable}
\usepackage{subcaption}

\AtBeginDocument{%
  \providecommand\BibTeX{{%
    \normalfont B\kern-0.5em{\scshape i\kern-0.25em b}\kern-0.8em\TeX}}}

\begin{document}

\title[A Legal and Empirical Analysis of the DSA Transparency Database]{Automated Transparency: A Legal and Empirical Analysis of the Digital Services Act Transparency Database}


\author{Rishabh Kaushal}
\affiliation{%
  \institution{Maastricht University}
  \country{Netherlands}}
\email{rishabh.kaushal@maastrichtuniversity.nl}
\affiliation{%
  \institution{Indira Gandhi Delhi Technical University for Women}
  \city{Delhi}
  \country{India}}
\email{rishabhkaushal@igdtuw.ac.in}

\author{Jacob van de Kerkhof}
\affiliation{%
  \institution{Utrecht University}
  \country{Netherlands}}
\email{j.j.w.vandekerkhof@uu.nl}

\author{Catalina Goanta}
\affiliation{%
  \institution{Utrecht University}
  \country{Netherlands}}
\email{e.c.goanta@uu.nl}

\author{Gerasimos Spanakis}
\affiliation{%
  \institution{Maastricht University}
  \country{Netherlands}}
\email{jerry.spanakis@maastrichtuniversity.nl}

\author{Adriana Iamnitchi}
\affiliation{%
  \institution{Maastricht University}
  \country{Netherlands}}
\email{a.iamnitchi@maastrichtuniversity.nl}
\renewcommand{\shortauthors}{R. Kaushal, J. van de Kerkhof, C. Goanta, G. Spanakis, A. Iamnitchi}

\acmYear{2024}\copyrightyear{2024}
\setcopyright{rightsretained}
\acmConference[ACM FAccT '24]{ACM Conference on Fairness, Accountability, and Transparency}{June 3--6, 2024}{Rio de Janeiro, Brazil}
\acmBooktitle{ACM Conference on Fairness, Accountability, and Transparency (ACM FAccT '24), June 3--6, 2024, Rio de Janeiro, Brazil}
\acmDOI{10.1145/3630106.3658960}
\acmISBN{979-8-4007-0450-5/24/06}


\begin{abstract}

The Digital Services Act (DSA) is a much awaited platforms liability reform in the European Union that was adopted on 1 November 2022 with the ambition to set a global example in terms of accountability and transparency. Among other obligations, the DSA emphasizes the need for online platforms to report on their content moderation decisions (`statements of reasons' - SoRs), which is a novel transparency mechanism we refer to as \textit{automated transparency} in this study. SoRs are currently made available in the DSA Transparency Database, launched by the European Commission in September 2023. The DSA Transparency Database marks a historical achievement in platform governance, and allows investigations about the actual transparency gains, both at structure level as well as at the level of platform compliance. This study aims to understand whether the Transparency Database helps the DSA to live up to its transparency promises. We use legal and empirical arguments to show that while there are some transparency gains, compliance remains problematic, as the current database structure allows for a lot of discretion from platforms in terms of transparency practices. In our empirical study, we analyze a representative sample of the Transparency Database (131m SoRs) submitted in November 2023, to characterise and evaluate platform content moderation practices. 
\end{abstract}

\ccsdesc[500]{Social and professional topics~Government technology policy}
\ccsdesc[500]{Information systems~Web applications}
\ccsdesc[500]{Human-centered computing~Collaborative and social computing design and evaluation methods}
\ccsdesc[500]{Applied computing~Law, social and behavioral sciences}

\keywords{Digital Services Act, Transparency, Computational Compliance.}


\maketitle

\section{Introduction}

The Digital Services Act \cite{DSA} is a much awaited platform regulation reform in the European Union that was adopted on 1 November 2022, with the ambition to set a global example in terms of accountability and transparency. Among other obligations, the DSA emphasizes the need for online platforms to report on their content moderation activities. In particular, Article 17 and 24 DSA govern so-called `statements of reasons' (SoRs) that are "clear and specific" explanations about content moderation decisions. Anonymized SoRs must be submitted to the Commission without undue delay, for their inclusion in "a publicly accessible machine-readable database managed by the (European) Commission" (Article 24(5) DSA). SoRs are currently made available in the DSA Transparency Database,\footnote{https://transparency.dsa.ec.europa.eu/} launched by the European Commission in September 2023. Very Large Online Platforms (VLOPs) already need to comply with these obligations (Article 92 DSA). These are online platforms that have at least 45 million monthly active users in the European Union. As of 17 February 2024, all online platforms in the European Union engaged in content moderation will be under an obligation to submit SoRs to the Commission. 

The DSA Transparency Database marks a historical achievement in platform governance. The database is a first of its kind, providing unprecedented insights into an otherwise entirely opaque world of content moderation \cite{Gorwa2020}, dominated by standards exclusively designed by platforms. This paper focuses on understanding whether the database delivers on the transparency objectives of the DSA with respect to SoRs as a novel computational transparency mechanism, which we call \textit{automated transparency} due to the fact that it is based on the standardization and automation required in providing large-scale data and information about content moderation decisions. To emphasize the practical role of automated transparency, we design two compliance-related questions, that we subsequently investigate through a multidisciplinary methodological approach using the legal doctrinal method for legal argumentation, and data science insights as evidence for the arguments made \cite{Smits2017}: 
\begin{enumerate}
    \item \textbf{RQ1}: To what extent does the Transparency Database structure (hereafter called `schema') as operationalized by the Commission, reflect the conditions of Article 17 DSA? 
    \item \textbf{RQ2}: What are the compliance issues of content moderation practices as reflected by VLOP submissions in the Transparency Database in relation to Articles 17 and 24(5) DSA? 
\end{enumerate}
 
In answering these questions, we make two main contributions. \textit{First}, we investigate and analyse the database schema, to better gauge what information VLOPs are required to provide, and how this information complies with the specific requirements of Articles 17 and 24 DSA. By comparing legal requirements with metadata categories (called `attributes' in the schema), we can identify the transparency gains but also potential limitations of the provided data, in terms of format and information depth. \textit{Second}, we empirically describe the information submitted through SoRs using exploratory data analysis techniques. We use a data sample (131m SoRs) reflecting all VLOPs in the Database at the time of retrieval, and formulate a series of data analysis tasks which provide further evidence relating to the early DSA-compliance strategies of VLOPs in their content moderation decisions.

\section{Prior work and contributions of the present study}

Launched on 25 September 2023,\footnote{https://digital-strategy.ec.europa.eu/en/news/digital-services-act-commission-launches-transparency-database.} the DSA Transparency Database is a relatively untapped resource. Previous work has focused predominantly on describing trends in content moderation on social media. Drolsbach and Prölloch~\cite{drolsbach_content_2023} used 156 million SoRs to describe content moderation practices for social media platforms over two months from the lens of data analysis, by looking at the different metadata reported in the Database. 
Dergacheva et al.~\cite{dergacheva_one_2023} use a different temporal scope in their study, analyzing 24 hours of SoRs for social media platforms (2 million SoRs). Their findings reflect those in~\cite{drolsbach_content_2023}, and provide additional analyses for content categories and the nature of illegal speech. Lastly, Trujillo et al.~\cite{trujillo_dsa_2023} analyse 195 million SoRs submitted by social media platforms on similar parameters (excluding those featuring free text), additionally providing cross-referencing from separate transparency reports. Similar to the other two studies, Trujillo et al. focus on social media platforms exclusively. Lastly, Kosters et al. \cite{kosters2024tiktok} performed a theoretical legal analysis of different levels of transparency obligations focusing on a single VLOP (TikTok).

Our study complements earlier works in three respects. First, previous empirical analyses have focused on SoRs from social media platforms only, as they have been the primary object of study in content moderation. Our contribution analyzes SoRs from \textit{all} VLOPs across different categories: \textit{shopping platforms} (Google Shopping, AliExpress, Zalando, Amazon), \textit{social media} (TikTok, Facebook, Pinterest, YouTube, Instagram, Snapchat, X, LinkedIn), \textit{app stores} (Google Play, App Store) and \textit{service platforms} (Google Maps, Booking.com). Second, our contribution is multidisciplinary, combining and contextualizing computational measurements with a legal dimension that critically reflects on the role of transparency in regulation. Finally, our descriptive exploratory analysis does not only focus on the pre-set attributes in SoRs, but we also provide further insights on the basis of linguistic data analysis on free text fields of DSA Transparency Database, to better understand platform submission practices and strategies. 

\section{Dataset}
Since its launch, the Transparency Database has been constantly updated in real time by VLOPs, who push data into the database at varying rates. Currently, all records are available for every day from the inception of the database. Because of the massive amount of data (as of January 20, the database contains 2,8 billion records), we focus on a ten-day window between 19 and 28 of November 2023. This dataset contains more than 131 million SoRs with an average of 12 million SoRs submitted each day. We note that this is an early snapshot of the database, as all online platforms in the EU (and not just VLOPs) will be required by law to start reporting in February 2024. 

Unlike other studies, we focused on a smaller time window to be able to analyse SORs submitted by all platforms, not only social media platforms. Our dataset comprises varying proportions (ranging between 6.2\% from Booking.com to 20\% from Facebook) of total SoRs submitted by each platform as of 15 December 2023. Since our focus is more on practices and database schema as they satisfy the DSA transparency objectives, this fraction of all SORs submitted so far is sufficient for our objective.  

\section{Statements of Reasons (SoRs) and DSA Compliance}

Transparency is commonly perceived to lead to good governance, based on a twofold assumption, the first being that transparency is the most direct way of achieving accountability, the second being that transparency allows regulators to better study the phenomenon they seek to regulate. Particularly in the context of the content moderation debate, this is pertinent; content moderation is notoriously opaque~\cite{pasquale_black_2016, Gorwa2020, Leerssen2023}, and calls for transparency have been common \cite{Roberts2018, suzor_what_2019}. In fact, Gillespie writes that "calls for greater transparency in the critique of social media are so common as to be nearly vacant"~\cite{gillespie_custodians_2018}. 
New in the pursuit of meaningful transparency in content moderation is a database containing content moderation decisions and statements of reasons of providers of online platforms. Article 24 DSA establishes transparency reporting obligations for providers of online platforms. In particular, paragraph 5 specifies that providers of online platforms must, without undue delay, submit content moderation decisions and statements of reasons to be included in a Commission-managed database. Although the DSA enters into force for the most part from 17 February 2024, it is applicable to VLOPs 4 months after they are first designated as VLOP (art. 92), meaning that currently 17 VLOPs are covered by the DSA. 
The Transparency Database, as proposed in Article 24(5), is revolutionary compared to other approaches such as transparency reports or audits. Although earlier efforts offered insights on an aggregate level about the scope of content moderation over periods of 6 months or more, the ins-and-outs of content moderation have been difficult to study since there was little transparency about individual decisions so far. The DSA calls the reporting of each individual content moderation decision as SoRs. This new transparency mechanism brings with it two novel features. First, it is a live database that needs to be updated regularly with ongoing content moderation decisions, reflecting the dynamic and hypercomplex ecosystem of content moderation. Second, it can be said to reflect raw data, and additional empirical analyses are not only possible but very much encouraged and facilitated by the Commission.\footnote{https://transparency.dsa.ec.europa.eu} To this extent, it can be characterised as an automated transparency mechanism that serves as another avenue for platform data access under the DSA.  

\subsection{SoRs requirements} 

SoRs submitted by the providers of online platforms have to adhere to the requirements laid down in Article 17 DSA.  Paragraph (1) requires platforms to report on restrictions imposed on users who post illegal content or content that is otherwise incompatible with their terms and conditions. 
Paragraph 3 lists the information that should be \textit{at a minimum} included in the SoRs:
\begin{itemize}
    \item \textit{A specification of the remedy.} Remedies may include removal or demotion of content, suspension or termination of account, suspension or termination of monetary payment, as well as the geographical and time parameters of these decisions. 
    \item \textit{Facts and circumstances}. Information about whether the decision was made based on a notification under Article 16 (including the identity of the notifier where strictly necessary), or on own-initiative investigations.
    \item \textit{Automated means of decision-making and detection}. This reflects whether the platform used automated means in detecting the content or making the content moderation decision.
    \item \textit{Legal grounds}. Where the content is deemed illegal, a reference needs to be made about the legal ground used to assess the content, and why the information is considered illegal. 
    \item \textit{Incompatibility with ToS}. Where the content is deemed to be incompatible with ToS, a reference needs to be made to the contractual ground and their interpretation. 
    \item \textit{Redress information}. This reflects the need to include clear and user-friendly information on redress possibilities to contest the decision, such as internal complaint-handling and out-of-court dispute settlement. 
\end{itemize}
Paragraph 4 establishes that the information in SoRs needs to be clear and easy to understand, but also precise and specific. 
SoRs must be uploaded without undue delay (Article 24(5) DSA), and must be free of any personal data. Failure to submit SoRs to the Transparency Database can be interpreted as non-compliance with the DSA, opening platforms to investigations and potential sanctions by national digital services coordinators (DSCs) from the Member State where the platforms are established (Article 51(2)), as well as from the Commission in the case of VLOPs (Article 73). 

\subsection{SoRs template and its compliance with the DSA} 

To implement the requirements in Article 17, the Commission has provided a template for SoRs which it expects platforms to submit via its designated API. This format operationalizes the conditions of Article 17 as `attributes', which can take two forms: \textit{free textual} (with character limitations sometimes applicable), and \textit{limited} (using pre-defined values from an array). The documentation also labels some of the attributes as mandatory three categories of attributes depending on whether they are mandatory, mandatory under certain conditions, or optional. The full SoRs schema, including the additional explanation for statement attributes is available in Section~\ref{sec:appendix}. In what follows, we inspect the API documentation\footnote{https://transparency.dsa.ec.europa.eu/page/api-documentation} and make a general overview of the standardized attributes requested by the Commission.  
In January 2024, the API documentation listed a total of 33 content-related attributes. Out of these, 23 feature limited input that needs to be selected from an array, and 10 feature free textual form. In the latter category, there are seven attributes limited to 500 characters (e.g. Legal Grounds Illegal Content), two limited to 2000 characters (e.g. Illegal Content Explanation), and one limited to 5000 characters (Facts and Circumstances Relied on in Taking the Decision). In turn, out of the total 33 attributes, eight are generally mandatory, 11 are mandatory under certain logical conditions relating to other categories, and eight are completely optional. In addition, for six attributes, the API documentation is unclear on mandatory nature. 

This brief investigation into the schema of SoRs gives us several important insights. 
First, it reveals whether the SoR standard as developed by the Commission is compliant with Article 17(3). This seems to be largely the case, with one notable exception: no reference is made to redress information, in spite of the fact that it is a DSA requirement (Article 17(3)(f) DSA).
Second, it reveals a balanced structure of the information SoRs are expected to deliver into the Transparency Database. However, the structure also raises further questions and concerns. Most of the fields are either conditionally mandatory, optional or unclear, which may be interpreted as a relatively high level of discretion from submitting platforms in terms of attribute choice. In addition, platforms can submit information in free textual form, and while this could potentially provide further information into platform-specific approaches, the only free textual attribute that is mandatory is `Facts and Circumstances Relied on in Taking the Decision'. The analysis in Section~\ref{sec:why} offers further insights into the practices and strategies undertaken by online platforms in submitting SoRs along the attributes discussed above.

\section{Platform Practices and early DSA compliance}

In this section, we report on the practices of the online platforms that have already submitted SoRs in the Transparency Database, and subsequently discuss the compliance implications of platform strategies in the early days of the database. Empirically, we offer evidence on multiple levels, roughly answering questions of who, how, why, what, when and where content moderation activity takes place. Specifically, we analyze \emph{who} reporting platforms/industries are  (Section~\ref{sec:who}); \emph{how} their moderation practices are performed, specifically whether in an automated way (Section~\ref{sec:how}); \emph{why} content is subjected to moderation actions (Section~\ref{sec:why}); \emph{what} the moderation actions taken are as reported to the database (Section~\ref{sec:what}); \emph{where} the content moderated comes from (in terms of language and legal territory) (Section~\ref{sec:where}); and finally, \emph{when} the moderation takes place compared to the time of content posting and the time of reporting to the database (Section~\ref{sec:when}). 

\subsection{Who is Reporting: VLOPs} 
\label{sec:who}

\begin{figure}[!h]
    \centering
    \includegraphics[width=0.5\textwidth]{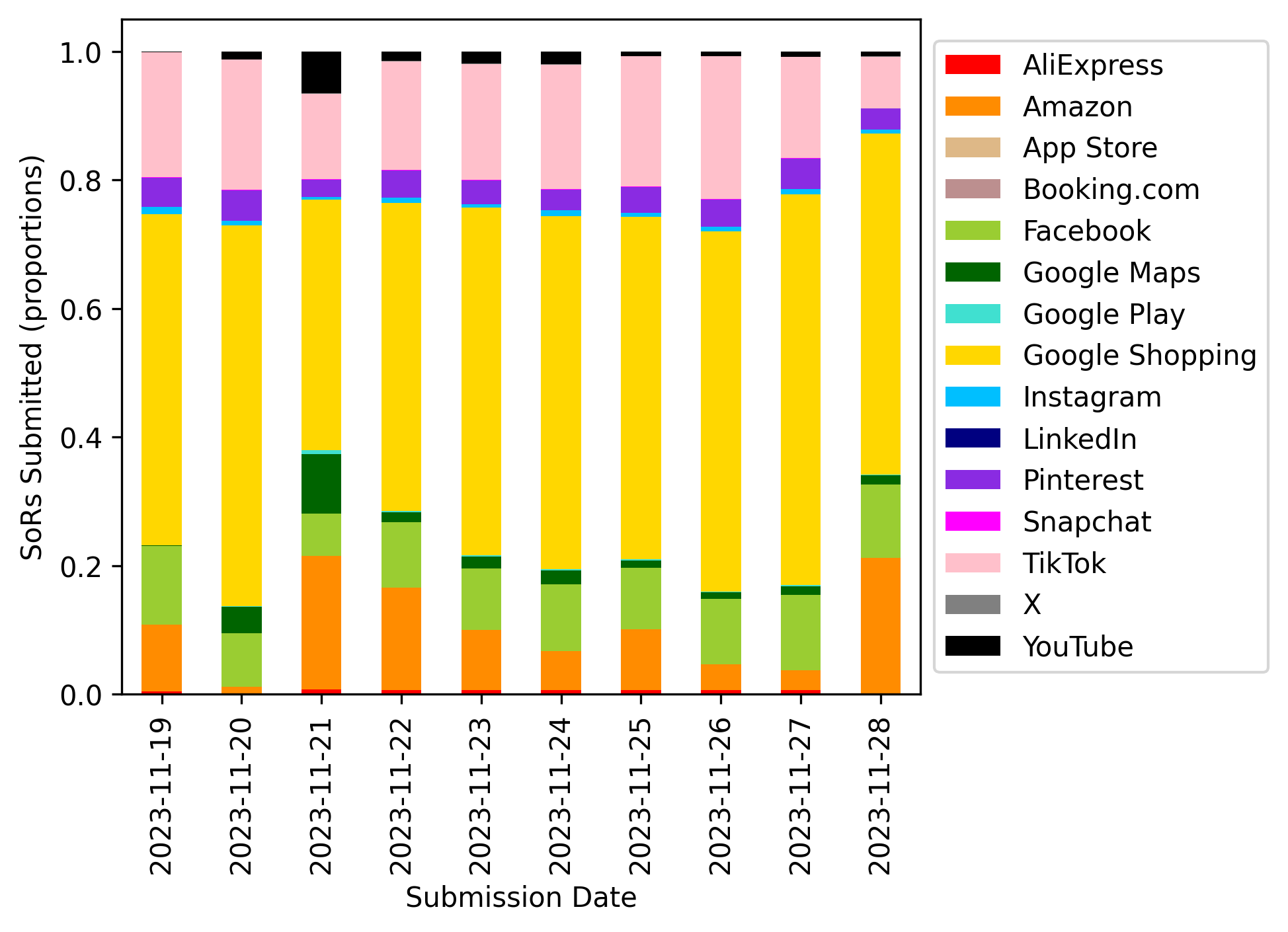}
        \caption{Fraction of SORs per day submitted by each VLOP.}
        \label{fig:plt_day}
\end{figure}

We investigate which platforms submit more data, and what the submission rate per platform is across the dataset. Figure~\ref{fig:plt_day} shows that Google Shopping has submitted more than half of the total SoRs (52.2\%), followed by TikTok (17.1\%), Amazon (10.8\%), Facebook (9.8\%), Pinterest (3.8\%), Google Maps (2.6\%), YouTube (1.8\%), Instagram (0.7\%), AliExpress (0.5\%), Google Play (0.2\%), Snapchat (0.08\%), X (0.03\%), App Store (0.005\%), Booking.com (0.004\%), and LinkedIn (0.003\%). Given that content moderation happens across all platforms, one would expect SoRs to be approximately correlated to the number of monthly active users on social media platforms in Europe. This does not appear to be the case. SoRs submission is dominated by Google Shopping, TikTok and Amazon, whereas these are not the largest VLOPs as reported to the Commission.\footnote{https://digital-strategy.ec.europa.eu/en/policies/list-designated-vlops-and-vloses.} This opens up many questions relating to the kind of content moderation activity done within these platforms and reported in the database. It also reflects the underestimated importance of content moderation on shopping platforms.

\textit{Compliance observations}. The Transparency Database provides an overview of platform practices in content moderation. However, it is difficult to estimate whether a low number of SoRs pushed in the Database can immediately translate to non-compliance. Still, in some cases, low SoR reports can be interpreted as a signal of questionable compliance in the context of additional transparency obligations.\footnote{See for instance the Commission's legal proceedings from December 2023 against X : https://ec.europa.eu/commission/presscorner/detail/en/ip\_23\_6709.}

\subsection{How Moderation Decisions and Detection are Done}
\label{sec:how}

\begin{figure}[!h]
        \centering
        \includegraphics[width=0.5\textwidth]{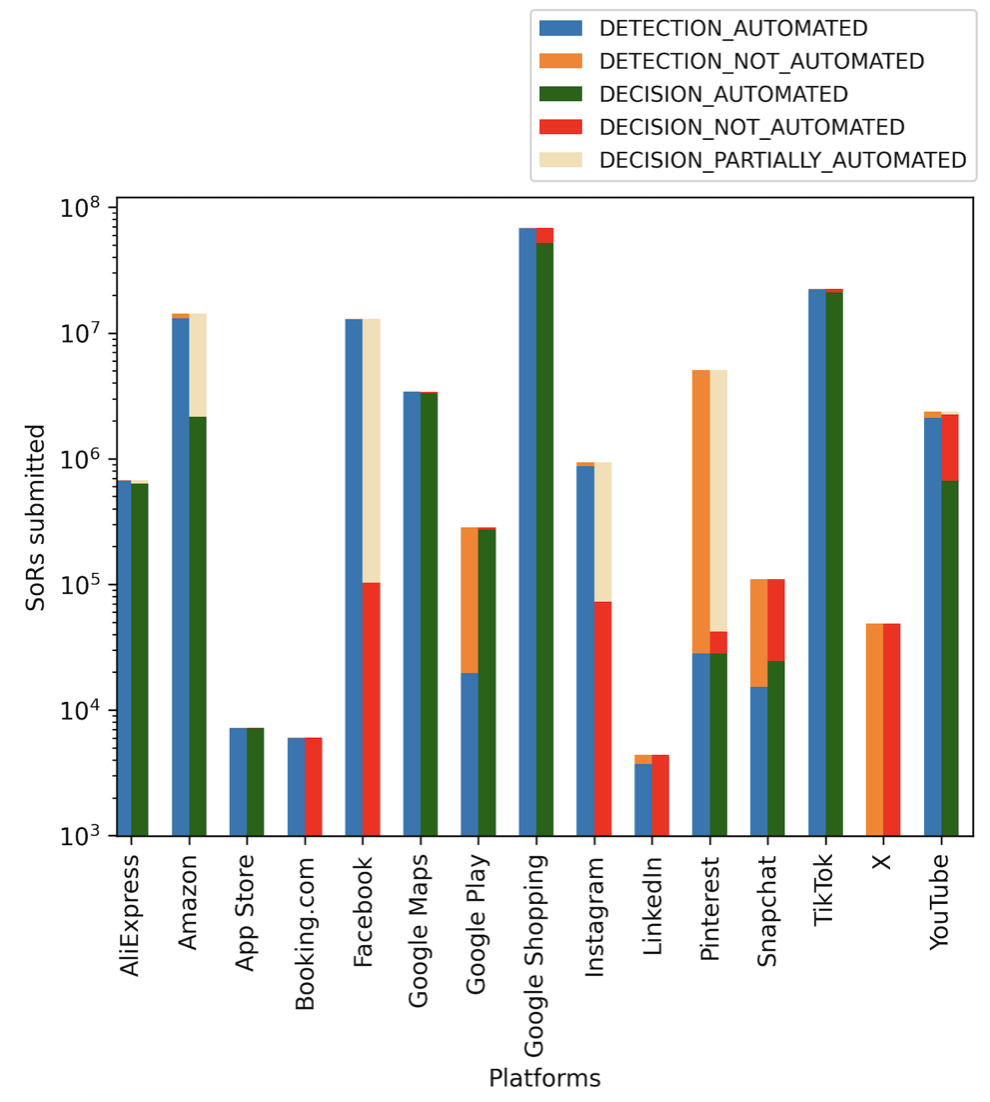}
        \caption{Distribution of $automated\_detection$ and $automated\_decision$ in SoRs.}
        \label{fig:auto_detect_decide}
    \end{figure}

Two attributes in the SoRs describe the manner of moderation: `automated detection' and `automated decision'. The `automated detection' field is binary and indicates whether automated tools were used to detect non-compliant content. Once content is detected as non-compliant, the decision on how to respond can be done in fully automated, partially automated, or in non automated manner, which is captured in `automated decision' attribute in SoRs. Figure~\ref{fig:auto_detect_decide} presents the distribution of automated detection and decision in our dataset. We find that X never reports automated means for detection, some platforms (such as Pinterest and Google Play) do not use automated means for detection in half of their SoRs, and others (such as App Store, Google Shopping and TikTok) rely mostly on automated detection. Regarding the decision for content moderation, X, LinkedIn and Booking.com do not automate at all, Facebook, Instagram, Snapchat, and YouTube make automated decisions for some SoRs, whereas AliExpress, Google Maps, Google Shopping, and TikTok make fully automated decisions for most (or all) SoRs. 
Also, we find that automated detection and fully automated decision are highly correlated in the moderation process, as evident in Figure~\ref{fig:auto_detect_decide}, where the height of bars corresponding to automated detection (blue) and automated decision (green) appear correlated. When platforms do not perform automated detection (orange), then they are more likely to perform partial automated decision rather than taking no automated decision or fully automated decision.

\textit{Compliance observations.} The DSA does not oblige platforms to moderate content automatically, in contrast to other legislation \cite{copyright}. However, it is generally considered desirable in the fight against harmful and illegal content that platforms undertake automated content moderation to cope with the immense volume of content shared online. This is visible in detection practices, which are predominantly automated across platforms, and partially in decision-making practices, which also involve a human touch. The involvement of humans is required by the DSA in cases of complaint procedures. Outliers who rely predominantly on human moderation are LinkedIn, Pinterest and X. The latter is currently under investigation for their content moderation practices, in which the lack of automated means can be used as a contraindication for proper moderation. 

\subsection{Why: Reasons for Moderation}
\label{sec:why}
Platforms may ground their content moderation decisions either on the law (illegal content) or their Terms of Service (ToS violations). Our dataset reveals that 99.8\% of the SoRs are reported as ToS violation and only 0.2\% as illegal content. 

 \textit{Illegal content.} For illegal content, two attributes (`illegal content legal ground', that includes up to 500 characters and  `illegal content explanation', up to 2000 characters) are specified in the Transparency Database for platforms to provide details of illegal content. Table~\ref{tab:illegal_dist} gives the distribution of the most prominent grounds (`illegal content legal ground') for marking content as illegal: AliExpress (66.2\%) tops the list, followed by X (18.7\%), Google Maps (8.3\%), YouTube (4.3\%), TikTok (2.0\%), Google Play (1.1\%), Facebook (0.07\%), and Instagram (0.02\%). Only 5 SoRs submitted by Google Shopping are flagged as illegal content. It appears that some VLOPs like Google Maps have mostly defamation related content whereas other VLOPs like TikTok have wide varying types of illegal content. We also notice that some VLOPs like Facebook specify only one standard reason \footnote{For example `This was a violation of sections 3.2 of our terms of service'.} for reporting all illegal content. We also find that the optional attribute `decision ground reference url' which is supposed to provide URL to the ToS or law, contains NULL values 99.9\% of the times. Having said that, there are additional attributes in the Transparency Database which give more details about the illegal and ToS violation content. Further, we looked at values of `illegal content explanation' which gives VLOPs more opportunity to give detailed explanation of grounds of flagging illegal content. To our surprise, only AliExpress makes use of this option. Out of 261,458 SoRs marked as illegal content, AliExpress gives specific and detailed explanations for 173,292 (66.2\%) of them, mostly in Chinese, followed by Spanish and French.  
 Besides AliExpress, YouTube (4\%) also provides explanations with userID and product name anonymized. 

\textit{ToS violations.} Google Shopping (52.2\%) submits more than half of the ToS violations followed by TikTok (17.1\%), Amazon (10.8\%), Facebook (9.8\%), Pinterest (3.8\%), Google Maps (2.6\%), YouTube (1.8\%), Instagram (0.7\%), AliExpress (0.3\%), and Google Play (0.2\%). The trend is completely different in terms of illegal content submissions. For ToS violation content, attributes namely `incompatible content ground' (upto 500 characters) and  `incompatible content explanation' (upto 2000 characters) are specified in the Database for VLOPs to use to provide details of ToS violation content. Table \ref{tab:tos_violation_dist} gives distribution of most prominent grounds (`incompatible content ground') for marking a ToS violation content by most active VLOPs submitting ToS violation content. As an illustration, Google Shopping does not provide any URLs to their ToS for the violations. However, they provide grounds for ToS violations, most of which is due to promotional overlays on images. Likewise, TikTok mentions harassment and bullying as the main grounds for ToS violations. 

\textit{Compliance observations.} The trend in the grounds for the content moderation decision is that such decisions are taken for violation of the platform ToS (99,8\% percent). This shows how platforms prefer to moderate of their community guidelines, rather than state law. Such a preference is not in itself non-compliant with the DSA, but it does create a paradox. The DSA places an emphasis on making what is illegal offline illegal online (recital 12). However, platforms do not moderate on what is considered `illegal', and therefore it can be argued the DSA does not bring too much added value to the existing moderation landscape. 

\begin{table*}
    \centering
    \begin{tabular}{p{2.5cm}|p{10cm}}
        VLOP & Illegal Content (Legal Grounds) \\
        \hline
        AliExpress & Illegal Content (67\%), Knowledge Platform Governance-General Infringement (30\%), Forbidden sale-prohibited ban on sale (1.6\%).\\
        X & NsfwAdultContent (47\%), NsfwViolenceGore (25\%).\\
        Google Maps & Defamation (99.9\%)\\
        YouTube & Copyright removal request (92.5\%), Privacy complaint (4\%). \\
        TikTok & Financial crime (47\%), Hate speech (13\%),  Other illegal content (9\%), Privacy violations (7\%), Non-consensual sharing of private or intimate images (6\%), Harrassment or threats (3.7\%), Terrorist offences/content (2.9\%), Child sexual exploitation (2.3\%), Defamation (2\%), Content relating to violent or organized crime (1\%).\\
        \hline
    \end{tabular}
    \caption{Distribution of grounds ($illegal\_content\_legal\_ground$) for top-5 VLOPs relying on illegal content.}
    \label{tab:illegal_dist}
\end{table*}

\begin{table*}
    \centering
    \begin{tabular}{p{2.5cm}|p{10cm}}
        VLOP & Ground of ToS Violation (Private Governance) \\
        \hline
        Google Shopping & Promotional overlay on image (37\%), Personalized advertising: Sexual interests (20.8\%),
        Personalized advertising: personal hardships (12\%), Restricted adult content (7.4\%), Adult-oriented content (3.6\%), Unsafe imagery (3.2\%).\\
        TikTok & Harassment and Bullying (32.9\%), Community Guidelines (26.1\%), Hate Speech and Hateful Behaviors (8\%), Sexual Exploitation and Gender-Based Violence (5.4\%), Dangerous Activities and Challenges (4.5\%), Alcohol, Tobacco, and Drugs (4.4\%), Sexually Suggestive Content (2.8\%).\\
        Amazon & Violation of section S1.1 of the Amazon Business Solutions Agreement (72.7\%), Violation of Amazons Listing Policies (9.5\%), Violation of the German Packaging Law (VerpackG) (4.7\%), Violation of the Electronical and Electronic Equipment Act (ElektroG – the German WEEE law) (3.1\%). \\
        Facebook & Violation of sections 3.2 of our terms of service (100\%).\\
        Pinterest & Adult content (87.4\%), Hateful activities (4.8\%), Graphic violence (2.7\%), Self-injury and harmful behavior (1.9\%), and 
        Medical misinformation (1.6\%).\\
        \hline
    \end{tabular}
    \caption{Distribution of grounds ($incompatible\_content\_ground$) for top-5 VLOPs relying on ToS violations.}
\label{tab:tos_violation_dist}
\end{table*}

In addition, the `decision facts' attribute is mandatory for SoRs, and it is central given that it describes the facts and circumstances relied on in taking the content moderation decision, which can help provide information about how platforms contextualize the complex rules they are bound to apply at scale. However, we find that this attribute mostly features generic templates that may not provide very meaningful information. Google Shopping, the most active VLOPs, submits only a standard template\footnote{`When reviewing content or accounts to determine whether they are illegal or violate our policies, we take various information into consideration when making a decision, including product data, website quality, merchant information, account information (e.g., past history of policy violations), and other information provided through reporting mechanisms (where applicable) and own-initiative reviews.'} in most (99.5\%) of the SoRs. The next key contributor of SoRs, TikTok, uses a similar albeit shorter statement\footnote{`The decision was taken pursuant to own-initiative investigations.'} in 98.5\% of its SoRs. AliExpress seems to be the only platform providing more specific details around moderation.\footnote{For instance, `The brand name of your product [id:abc] has been selected incorrectly and therefore cannot be published'; or `To xyz: Your product [id: abc] is similar to the def logo and important product elements, which can easily cause confusion among consumers, so it cannot be published'.} 

\textit{Compliance observations.} Article 17(1) requires SoRs to be sufficiently clear and specific. This allows users to be informed about the restrictions imposed on their content or accounts, to seek redress, and potentially alter their behaviour. SoRs can include up to 5000 characters describing facts and circumstances, and yet most such sections seem to be automatically generated and overly short. This does not necessarily allow users to identify what elements of their content are violating norms leading to restriction, nor does it allow them to form a well-argued case for redress, which remains a large problem for content moderation. Our analysis raises the specific question of whether generically formulated facts and circumstances comply with requirements of clarity and specificity. Since this is a novel question for compliance, additional research is needed to offer a more concrete answer.

\subsection{How Platforms Sanction: What Moderation Actions are Reported}
\label{sec:what}
Content moderation has long been considered in binary terms, namely with content either staying up or being taken down. However, content moderation reflects a wide array of ways in which platforms can sanction undesirable content and user behavior \cite{Leerssen2023}. Most of the restrictions are done at the level of content visibility (93.4\%), followed by account (6.5\%), service provision (1.1\%), and monetary limitations (0.015\%). Every SoR has at least one of these four types of restriction. In terms of subcategories, \textit{restriction by visibility} is further distributed among values of disabled content (66\%), removed content (24\%), and other (8\%). Very rarely, content has been demoted (0.08\%) and age restricted (0.03\%).  Content is mainly disabled by Google Shopping (84\%), followed by Amazon (14\%). Most of the content removal is done by TikTok (56\%), Facebook (15\%), Google Maps (12\%), Amazon (8\%), YouTube (7\%), and Pinterest (4\%). Demotion of content is only being done Facebook (98.2\%) and Instagram (1.8\%) whereas age restricted of content is only being done by TikTok (90.3\%) and YouTube (9.7\%). The most common reasons for `others' are `Video not eligible for recommendation in the For You feed' (50.4\%), `Limited distribution' (38.14\%), `LIVE not eligible for recommendation and restricted in search results for 10 minutes' (7.9\%), and `LIVE not eligible for recommendation and restricted in search results (1.9\%). For \textit{account restriction}, 89.9\% of the times the account is suspended, and termination accounts for 10\% of the times. Most of the account suspension is done by Facebook (88.9\%) followed by Google Shopping (4\%), Snapchat (2\%) and Amazon (2\%). Most of the account termination is done by Instagram (6.2\%) and TikTok (3\%). Most of the \textit{service restriction} result in partial suspension (84\%), total termination (8.7\%), total suspension (6\%), and partial termination (0.3\%). Most of the partial suspension is being done by Amazon (76\%) and TikTok (7\%); total suspension by Amazon (5.7\%); and total termination by Google Maps (5.5\%). All of the \textit{monetary restriction} results into suspension, and all of this is contributed by Amazon. No other VLOP is performing restrictions at monetary level.

\textit{Compliance observations.} The DSA does not pose strict requirements as to the type of restriction that needs to be imposed when content violates terms of service or state law. However, recital 77 stresses that, for VLOPs, unnecessary restrictions on the use of their service need to be avoided, in light of the potential fundamental rights risks that such restrictions pose. Platforms may therefore be inclined to seek restrictions that are proportional to the violation, and prefer less impactful restrictions such as demotion or demonetization as opposed to blocking account or content indefinitely where possible. This is not necessarily reflected in the dataset, but it is difficult to argue that such would be non-compliant with the DSA. Fundamental rights considerations are considerably more important for social media platforms than for other categories, such as marketplaces or app stores. From this perspective, sanctions might have a different impact. Suspending the monetization of a seller account on Amazon will be considerably different than suspending the monetization of a Live stream on TikTok. These implications need further elaboration in future research. Yet one observation is striking. Literature on platform governance has increasingly emphasized demonetization as a content moderation wrong by social media platforms \cite{Caplan2020}. Our dataset does not show any such demonetization taking place in the EU. Recital 55 refers to monetisation as the advertising revenue-driven provision of information, but the true scope of this concept remains unclear. This may be interpreted as social media platforms not yet reporting on their full monetization spectrum.

\subsection{Where Content is Moderated: Language and Territorial Scope}
\label{sec:where}

Language is often a proxy indicator of the region from where the content was posted, or of a platform's operational language in a given geographical area. In this section, we offer some language-based insights for SoRs, and we also revisit some earlier observations in the light of these insights. The optional attribute `content language' in the SoR schema is only populated in approximately 50\% of the records in our dataset. German (DE, 34.2\%) is the most frequent language, followed by English (EN, 26.5\%), French (FR, 9.9\%), Italian (IT, 9.3\%), Spanish (SP, 3.3\%), Polish (PL, 2.9\%), Dutch (NL, 2.7\%), Czech (CS, 2\%), Romanian (RO, 1.7\%), and Slovak (SK, 1.4\$). Only three VLOPs, namely, Google Shopping, Google Play, and Booking.com, fill the language of content.  

Additionally, SoRs feature several text fields that can be used as a more accurate proxy of the language of the moderated content. Overall, we inspect the language used in the following fields that require platforms to enter text in free form (with some limit in the number of characters): (a) \textit{Facts and circumstances relied on in taking the decision} (decision\_facts) (limit of 2000 characters, (b) \textit{Illegal Content Legal Grounds} and \textit{Illegal Content Explanation} are two fields that are required in a SoR if `decision ground' was illegal content, (c) \textit{Incompatible Content Grounds} and \textit{Incompatible Content Explanation} are required if `decision ground' was incompatible content.
For the decision\_facts, all platforms with the exception of AliExpress (where the dominant language is Chinese) report content in English. Notably, there is a difference in the length of the description with the shortest facts being provided by Instagram and Facebook (51 and 55 characters respectively) and X (59 characters) and the longest facts being provided by Google Maps/Play/Shopping (438/468/433 characters respectively) and YouTube (635 characters). These numbers also correlate with the number of unique entries by platforms in this field, e.g. Instagram and Facebook provide only 2 entries here, namely "This content was determined to violate terms of use" and "This account was determined to violate terms of use".
For the Incompatible Content Grounds, all platforms except AliExpress and TikTok post content only in English (in total 13,0861,024 SoRs). For AliExpress the dominant language is Chinese (>99\% of cases), while TikTok provides grounds in 31 languages. More specifically, in the cases of booking.com, Facebook, and Instagram there is only one standard entry for all relevant SoRs: Violation of article 1F in section How we work in Terms and Condition, This was a violation of "sections 3.2" of our terms of service and This was a violation of  "How You Can`t Use Instagram section" respectively. Notably, for the Incompatible Content Explanation and AliExpress, there is some incompatibility between the language of the grounds (Chinese) and of the explanation (could be Spanish). 

For the Illegal Content Grounds 10 platforms have logged 261,458 SoRs (AliExpress, Facebook, Google Maps, Google Play, Google Shopping, Instagram, Snapchat, TikTok, X, YouTube). All platforms post in English except AliExpress where Chinese is the dominant language in almost all entries. Facebook, Instagram, Snapchat provided only one entry for their relevant SoRs, namely "This was a violation of "sections 3.2" of our terms of service", "This was a violation of  "How You Can`t Use Instagram section" of our terms of use" and "Illegal or regulated activities" respectively. Other platforms like YouTube provide more extensive information (e.g. Hate Speech Law - FR or Violent Extremism Law - GR). For the Illegal Content Explanation, AliExpress has similar language patterns as with the Incompatible Content Explanation. All other platforms use English (except for very few TikTok examples TikTok which are in French).

A different location-based attribute in the SoRs is `territorial scope', which records the territorial scope of the restrictions imposed - in other words, the geographical limitations applied to content visibility, which we could also refer as the geo-blocking of content as a moderation sanction. 72.68\% of the SoRs apply territorial limitations to the entire European Economic Area (EEA), followed by 16.34\% to the entire European Union. There are also a few specific national level restrictions applied to Germany, France, Spain, and Italy, respectively. 

\textit{Compliance observations.} The DSA is silent about any language requirements relating to the submission of SoRs. In principle, it can be argued that SoRs could be submitted in any of the 24 official languages of the European Union. Following this perspective, submitting SoRs in Chinese could pose compliance questions. However, more important questions emerge when gauging the scope of the language and territoriality dimensions of the Transparency Database. The whole rationale of the Database is to provide a new, standardized tool for content moderation transparency. As we discovered, some SoRs are even submitted in multiple languages, and overall, the languages used vary significantly. This poses considerable computational constraints for the further verification of the submitted information, and future research should also be geared towards unveiling what that means in practice for the computational burden placed by platforms on public authorities and other parties interested in the DSA Transparency Database. 

\subsection{When SoRs are Submitted (``Without Undue Delay'')}
\label{sec:when}
The DSA requires SoRs to be submitted without undue delay to the transparency database. 'Without Undue Delay' is a common phrasing in the DSA for a variety of platform obligations, but is not specified in this particular context. To understand the delays incurred in content moderation, we propose two metrics: moderation delay and upload delay. We define \textit{moderation delay} as the delay between when the content was originally posted on platform and the time when content was moderated. We define \textit{upload delay} as the interval between the time when the content was moderated by the platform and the time when its corresponding SoR was uploaded in the Transparency Database. Note that all three time records needed for this are reported in a SoR. 

\begin{figure*}
\centering
\begin{subfigure}{.5\textwidth}
  \centering
  \includegraphics[width=.9\linewidth]{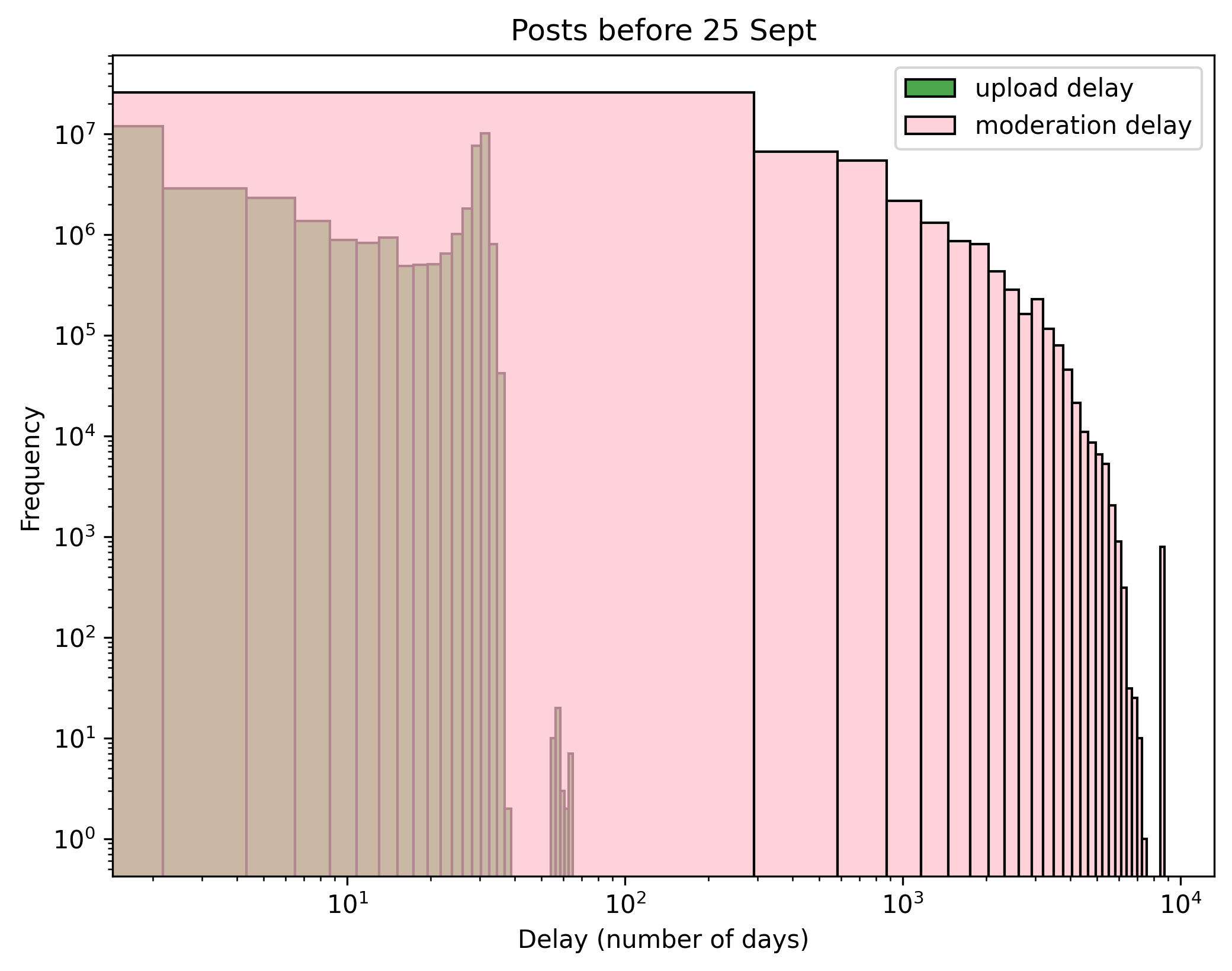}
  \caption{before 25 Sept}
  \label{fig:cmp_before_delay}
\end{subfigure}%
\begin{subfigure}{.5\textwidth}
  \centering
  \includegraphics[width=.9\linewidth]{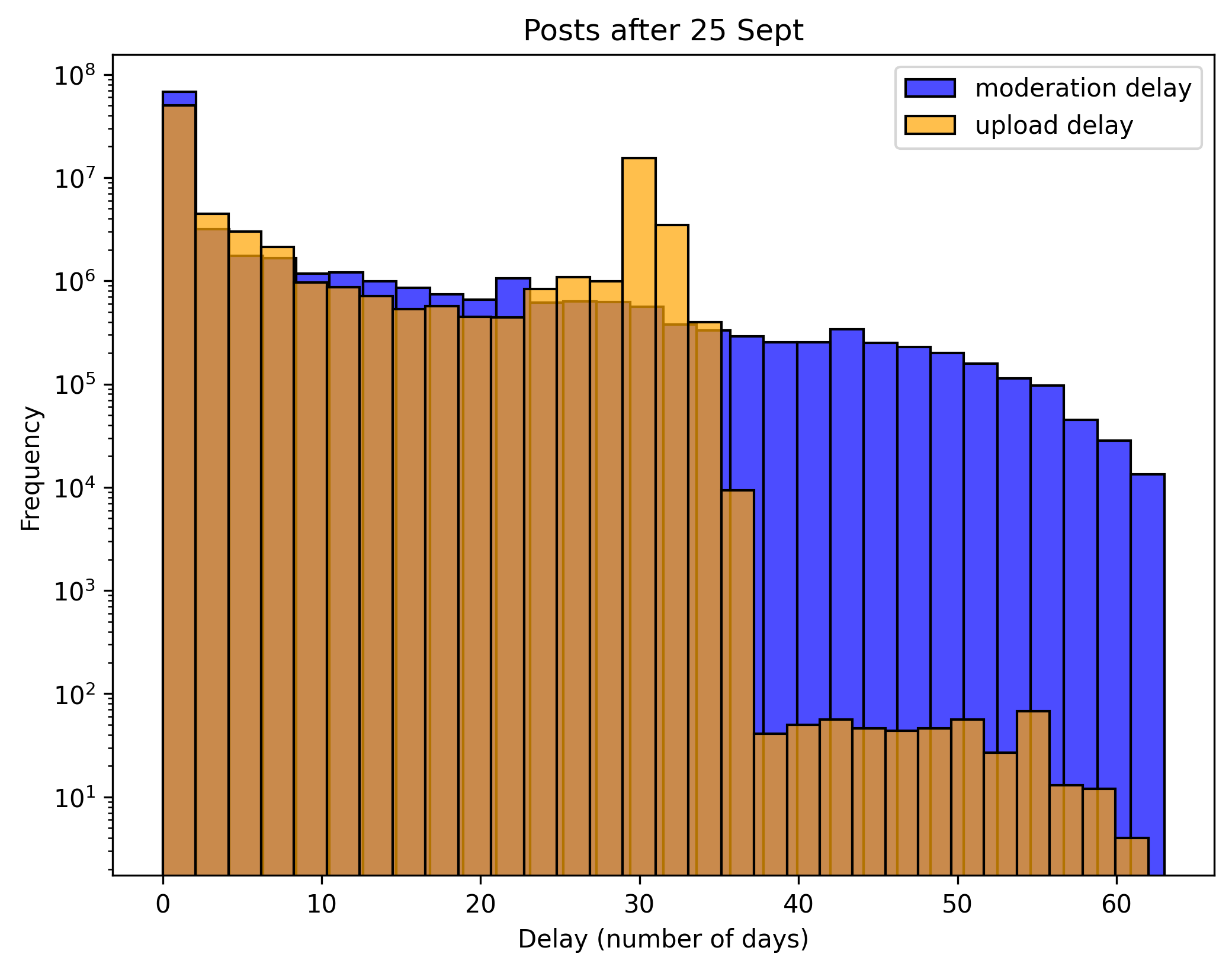}
  \caption{after 25 Sept}
  \label{fig:cmp_after_delay}
\end{subfigure}
\caption{Moderation and upload delay for content posted before and after the launch of the Transparency database.}
\label{fig:cmp_delays}
\end{figure*}

In theory, the DSA became applicable for VLOPs four months after the Commission notified the platforms it established to have VLOP designation (Article 33(6) DSA). The Commission made this designation public on 25 April 2023.\footnote{https://ec.europa.eu/commission/presscorner/detail/en/ip\_23\_2413} According to this timeline, VLOPs should have already submitted SoRs as of end of August. However, the Transparency Database was only launched on 25 September 2023, so we will consider this as the actual moment from when compliance can be considered. Once the database was launched, VLOPs submitted SoRs that predate the starting of the DSA. Overall, from the SoRs submitted on content that was posted prior to 25 September 2023, the mean moderation delay is 457.9 days (with a standard deviation of 599.6 days) while for content posted after, the mean moderation delay as reported is only 3.9 days (standard deviation 9.5 days). This significant speeding up in monitoring practices may be due to a variety of reasons, from maturing moderation practices within the company to preparing for complying with the DSA. The upload delay for content posted after the launch of the database is on average 9.6 days (with a standard deviation of 12.7 days). Details on the frequency of such delays as present in our dataset are shown in  Figure~\ref{fig:cmp_delays}. It appears that after the launch of the database, most of the content is moderated within a few days of posting. 

\begin{figure*}
\centering
\begin{subfigure}{.5\textwidth}
  \centering
  \includegraphics[width=.9\linewidth]{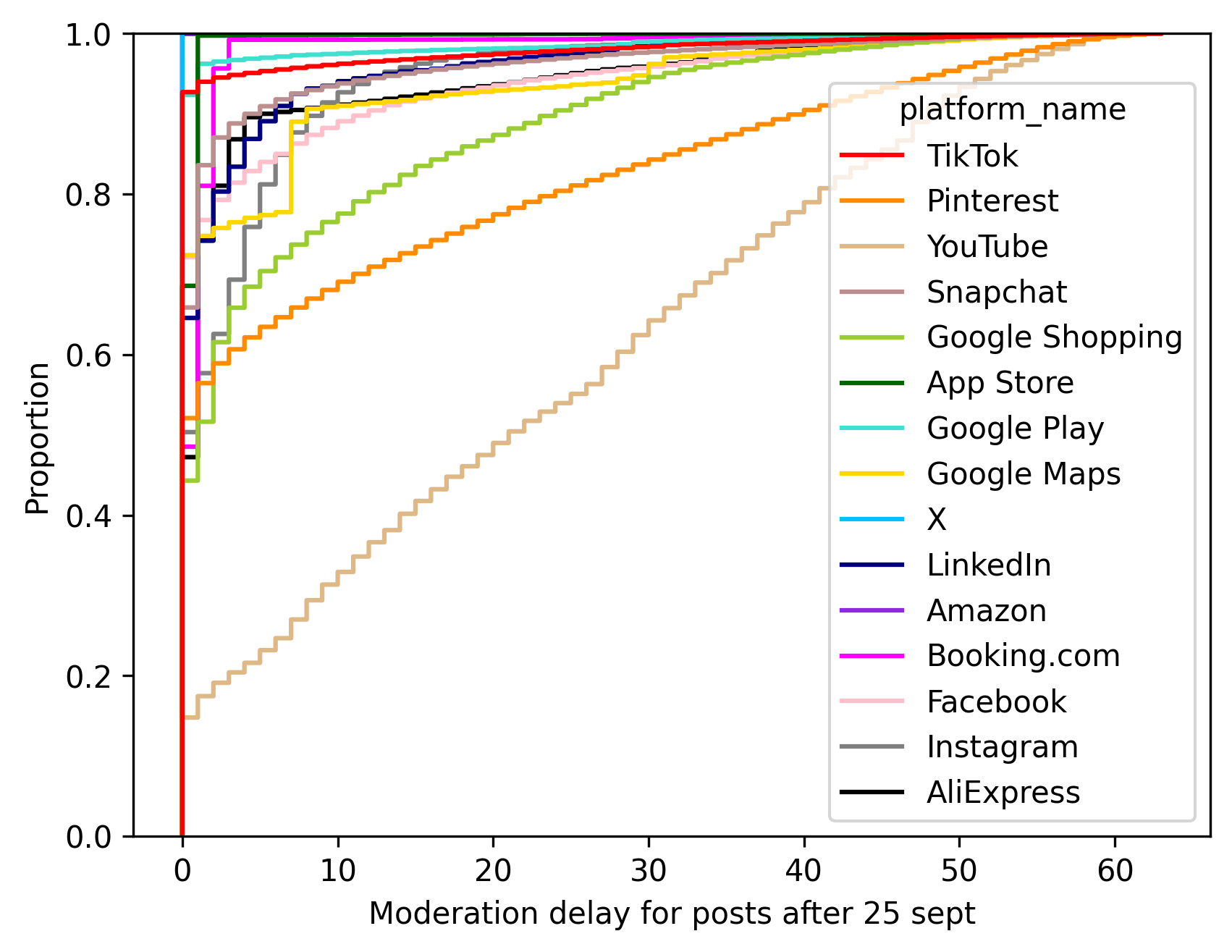}
  \caption{Moderation delay}
  \label{fig:cdf-moderation-after}
\end{subfigure}%
\begin{subfigure}{.5\textwidth}
  \centering
  \includegraphics[width=.9\linewidth]{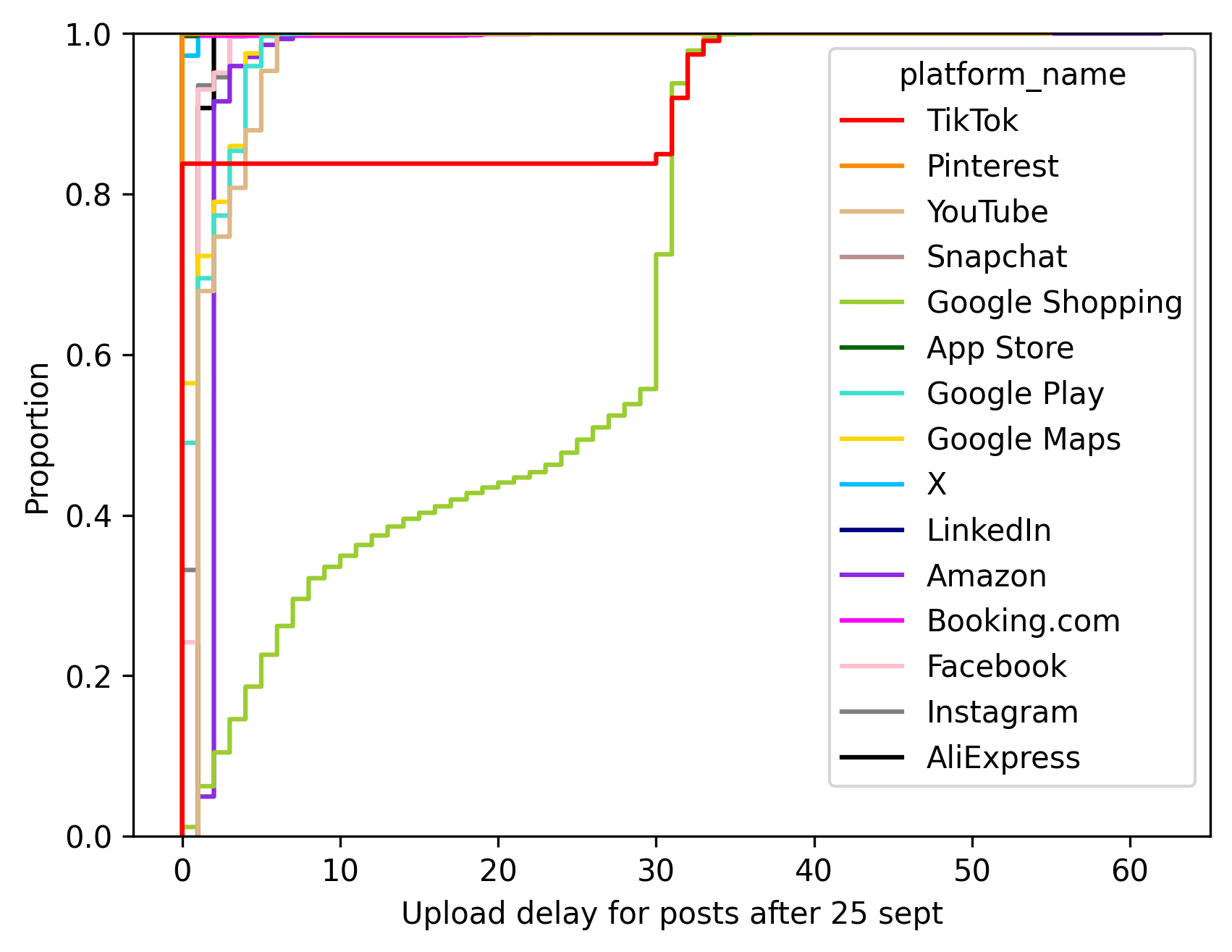}
  \caption{Upload delay}
  \label{fig:cdf-upload-after}
  \end{subfigure}%
\caption{Cumulative distribution function of the moderation delay (time between content posting and moderation decision) and upload delay (time between moderation decision and reporting to the Transparency Database) of content posted after the database launch.}
\label{fig:cdf-delays}
\end{figure*}

Figure~\ref{fig:cdf-delays} presents cumulative distribution functions on the moderation and upload delay per platform for content posted after the launch of the transparency database. We note that Google Shopping is slower in reporting its SoRs than it is in moderating its content, which might be due to the enormous amount of reporting it has to do. (Google Shopping reports the majority of the SoRs in the database, possibly stressing the infrastructure). 

\textit{Compliance observations.} Article 6 DSA requires platforms to remove illegal 'expeditiously' upon learning of its existence in order to avoid liability for that content. What timeframe is meant by 'expeditiously' is undefined, however. Some practices revolved around deadlines such as 24 hours or a week. That said, this is highly dependent on the nature of the content and the circumstances of the case. It is therefore difficult to ascertain whether there is non-compliance. What is visible is that platforms that detect illegal content manually, such as Pinterest, experience \textit{moderation delay.} Platforms that detect illegal automatically are less likely to experience \textit{moderation delay}, with the exception of YouTube. However, it is dependent on the nature of the content restricted whether this is non-compliant with the DSA.
Next to that, Article 24(5) requires platforms to submit SoRs to the database without undue delay. It is presumed that the period for submission starts when the SoR is submitted to the recipient of the service. Therefore, platforms need to avoid \textit{upload delay}. The delay varies per platform, with the upload delay being dependent on whether the platform automates that upload or does so manually, or in batches. Compliance is ensured by automating the submission process, leading to automated transparency. Since there are no norms defined in the DSA, it is difficult to ascertain whether there is non-compliance.

\section{Conclusion: Does the DSA's automated transparency rise to its promise?}

Our paper focused on investigating the DSA through the lens of a novel transparency mechanism aimed to offer insights into the opaque world of content moderation \cite{Gorwa2020, MyersWest2018, Roberts2018}. We referred to the DSA Transparency Database as a form of \textit{automated transparency}, because it is based on a standardized process through which platforms are supposed to share structured data about their content moderation decisions. The Transparency Database is an ambitious project with no precedent. A certain degree of voluntary transparency has been commonplace for content moderation since around 2010, when Google started disclosing information on its Government Requests,\footnote{https://transparencyreport.google.com/government-removals/overview?hl=en.} followed by Twitter in 2012.\footnote{https://transparency.twitter.com/en/reports/removal-requests.html\#2021-jul-dec.} Presently, most social media platforms endorse transparency in their content moderation practices, signing self-regulatory instruments requiring transparency such as the Santa Clara Principles on Transparency and Accountability in Content Moderation \cite{prerna_2020}, as well as co-regulatory instruments such as the EU Code of Conduct against Illegal Hate Speech \cite{EU_code} and the EU Code of Practice against Disinformation \cite{EU_code_22}. Regulators have also taken to the idea of transparency reporting, resulting e.g. in bi-annual transparency reporting under Germany's stringent platform liability regime \cite{heldt_reading_2019}. Yet transparency should play a meaningful role in establishing accountability for the purpose of user benefits \cite{Shagun_2019, suzor_what_2019, fung_full_2007}. Transparency can be relatively meaningless if mainly used to control public discourse \cite{flyverbom_transparency_2016, thevergeAirbnbsWorst}. 

Overall, we find the DSA Transparency Database to be a fascinating resource of data and information on content moderation, which aligns with earlier scholarly calls of automating enforcement in the digital sector \cite{goanta_2022}. While by no means a perfect tool, the Transparency Database has helped reveal the sheer scale of content moderation activities, and is providing public data relating to the content moderation process in ways that so far were not accessible for civil society, researchers or public administration. In addition, the Transparency Database emphasizes the need to study content moderation from a broader perspective. So far, the content moderation debate has mostly focused on social media platforms, but the Database shows how important content moderation is for other consumer markets such as app stores, service platforms and marketplaces. Going further, content moderation studies must acknowledge this broad nature, particularly in the light of the DSA's generous definition of illegal content. By looking at the compliance of various features of the DSA Transparency Database with the DSA itself, we set out to understand how automated transparency looks like in practice in order to critically reflect on its usefulness in scrutinizing platforms and making better policy. We conclude our study with three discussion points. 

Firstly, it is important to realise that the Database only provides insight in SoRs submitted by platforms, and therefore content moderation can currently only be studied to the extent that platforms allow it to be studied. In their work, Trujillo et al.~\cite{trujillo_dsa_2023} already found inconsistencies between the Transparency Database and the DSA-mandated Transparency Reports, showing the importance of having multiple transparency mechanisms. This has been the case earlier with various disclosure combinations such as court-mandated data sharing by platforms and voluntary transparency reports \cite{thevergeAirbnbsWorst}. Time and time again, it has been shown that over-relying on platform power for data access leads to data manipulation risks that may defeat the entire purpose of transparently sharing the data in the first place \cite{thevergeFacebookReportedly}. Apart from complementary transparency mechanisms, the investigation and enforcement powers of the European Commission can also play a role in ensuring further accountability. Article 40 of the DSA, often referred to in the context of access to data by vetted researchers \cite{edelson_23, Nonnecke2022}, is also the locus of data access powers by public authorities tasked with the enforcement of the DSA. The interplay between all possible platform insights will require further analysis and discussion. A key take-away from empirically studying the DSA Transparency Database is that platforms can strategize their use of the database as a means to show their compliance, even though in practice it may be insufficient. 

Secondly, the Database creates the possibility to remain opaque on the grounds behind content moderation decisions. The DSA has attempted to make 'what is illegal offline, illegal online', reflected in recital 12 and Article 3(h). This is a laudable standpoint from a legal certainty perspective aimed to align private governance with legal standards. However, as we have seen in our analysis, most content moderation decisions in the Transparency Database (>99,8\%) are based on ToS infringements. On the one hand, this can be interpreted as platforms shirking away from any responsibility of interpreting the law: platforms are not state regulators, and nor are they judges. On the other hand, however, since ToS are private agreements between user and platform that may unilaterally be changed at any moment in time, this creates a risk for legal certainty in terms of the forseeability of what is considered illegal online. This makes it difficult to discuss trends in illegal content online based on the Database, and therefore may not contribute to meaningful transparency. 

Lastly, the Transparency Database is built on a standardization process undertaken by the European Commission. It is true that considerable efforts and resources have gone into setting up the enforcement framework of the DSA. The Commission is visibly increasing its legal and technological capacities in the process.\footnote{See for instance https://digital-strategy.ec.europa.eu/en/news/do-you-want-help-enforce-digital-services-act-apply-now-be-part-dsa-enforcement-team.} It is however debatable if these efforts are enough to bridge the considerable power and technology gap that currently exists at any level of governance between the public sector and VLOPs in particular. This knowledge asymmetry will inevitably affect what and how the Commission and national DSA enforcers will ask from online platforms. We already saw in our analysis that platforms are required to contextualize content moderation by providing original, free text insights into their practices, only for them to standardize basic 50 character sentences that do not add any value to a SoR. As we have explored in the compliance observations throughout Section 5, it is currently difficult to ascertain if many of these practices are actually compliant with the DSA. This uncertainty may weaken the Commission's negotiation stance with VLOPs in particular for further alignment with the objectives of the DSA

Automated transparency mechanisms such as the DSA Transparency Database can certainly benefit from additional multidisciplinary attention. Future research should also consider comparisons between VLOPs and other online platforms which will also be under an obligation to submit SoRs as of 17 February 2024.

\begin{acks}
CG is supported by the ERC Starting Grant research project HUMANads (ERC-2021-StG No 101041824) and the Spinoza grant of the Dutch Research Council (NWO), awarded in 2021 to José van Dijck, Professor of Media and Digital Society at Utrecht University. 
\end{acks}

\bibliographystyle{ACM-Reference-Format}
\bibliography{main}

\newpage

\section{Appendices}
\label{sec:appendix}

\subsection{SoR schema design}

We provide the schema design of SoR in the Table \ref{tab:sor_design}.
\clearpage
\onecolumn
    \begin{longtable}{p{.36\textwidth} | p{.64\textwidth}} 
    \caption{Schema of SoR designed by Commission as per Article 17 DSA. Fields marked * are mandatory.} \\
        \toprule
        Attribute & Description \\
        \hline
        $uuid^*$ & unique identifier for each SoR. \\
        \hline
        \multicolumn{2}{l} {Content related attributes} \\
        \hline
        $content\_type^*$ & type (text, image, video, etc) of content. \\
        $content\_type\_other$ & more details when content type is OTHER. \\
        $content\_language$ & language of the content. \\
        $content\_date^*$ & date when content was posted on VLOP.\\
        \hline
        \multicolumn{2}{l} {Type of restriction (visibility, monetary, provision of service, and account)} \\
        \multicolumn{2}{l} {Any one of the fields marked with \# has to be filled.}\\
        \hline
        $application\_date$ & date when moderation decision was made by the VLOP. \\
        $decision\_visibility^\#$ & restriction w.r.t. visibility can be categorized based on whether content is removed or disabled or age-restricted or others. \\
        $decision\_visibility\_other $ & to be filled when decision visibility doesn't fall into any pre-defined types like `limited distribution', `Video not eligible for recommendation in user's feed', etc.\\
        $end\_date\_visibility\_decision$ & date when visibility restriction ends for the flagged content. \\
        $decision\_monetary^\#$ & restriction (suspension or termination) w.r.t. monetary payments associated with the content being flagged.\\
        $decision\_monetary\_other $ & to be filled if other type of monetary restriction is being applied by VLOP which is not covered in pre-defined types. \\
        $end\_date\_monetary\_restriction$ & date when monetary restriction ends for the flagged content.\\
        $decision\_provision^\#$ & restriction w.r.t. provision of service (partial or total suspension, partial or total termination of service) to the recipient of service.\\
        $end\_date\_service\_restriction$ & date when provision restriction ends for the flagged content.\\
        $decision\_account^\#$ & restriction w.r.t. account being suspended or terminated. \\
        $end\_date\_account\_restriction$ & date when account restriction ends for the flagged account.\\
        $account\_type$ & type (business or private) of account. \\
        $territorial\_scope^*$ & it specifies the territorial scope of restriction.\\
        \hline
        \multicolumn{2}{l} {Grounds on which decision is made, either incompatible or illegal content} \\
        \hline
        $decision\_ground^*
        $ & reason for decision whether incompatible (if post is against terms of usage of a particular VLOP) or illegal content (if post is in violation of a national law). \\
        $decision\_grounds\_reference\_url$ & link to the law or ToS of VLOP that is being violated. \\
        $incompatible\_content\_grounds (text)$ & reasons for content being incompatible which could include harassment \& bullying, community guidelines, adult content, hate speech, violation of Terms of Service of VLOP, etc. \\
        $incompatible\_content\_explanation (text)$ & detailed explanations for reasons that content is incompatible. \\
        $illegal\_content\_legal\_ground(text)$ & content is flagged illegal because it violates a national law.\\
        $illegal\_content\_explanation (text)$ & grounds of marking content illegal. \\
        $incompatible\_content\_illegal$ & it indicates whether moderated content was incompatible but also illegal.\\
        $category^*$ & specifies the category of decision whether it is related to Terms of Service, illegal or harmful content, pornography, violence, unsafe products, and so on. \\
        $category\_addition$ & any other additional category other than pre-defined list of categories. \\
        $category\_specification$ & it specifies more keywords related to the category attribute.\\
        $category\_specification\_other$ & it is filled when KEYWORD\_OTHER is specified in category specification. \\
        $decision\_facts^* (text)$ & it is a text field that gives details of the circumstances and facts relied upon in taking moderation decision.\\
        $source\_type^*$ & it specifies a pre-defined set of types of circumstances relied upon in taking moderation decision.\\
        $source\_identity$ & it is optional field specifying notifier or source which helped in making moderation decision.\\
        $automated\_detection^*$ & it specifies whether automated means (yes or no) were used to make the detection. \\
        $automated\_decision^*$ & it specifies the extent to which (fully or partially automated or human intervention) automated means were used to derive at the decision. \\
        $created\_at$ & date when SoR was submitted to the DTD database.\\
        \bottomrule
    \label{tab:sor_design}
\end{longtable}
\clearpage
\twocolumn

\end{document}